# HOLE SUPERCONDUCTIVITY IN $MgB_2$, CUPRATES, AND OTHER MATERIALS

J. E. Hirsch

Department of Physics, University of California, San Diego,
La Jolla, CA 92093-0319

## I. INTRODUCTION

Fifteen years after the discovery of high $T_c$ superconductivity in cuprates, the condensed matter community has again been jolted by an unexpected discovery, 40K superconductivity in $MgB_2$ [1]. During the past 15 years, theorists have struggled to understand the many apparently non-conventional features of superconductivity in the cuprates, however so far no fundamental and generally agreed upon understanding of the mechanism leading to their superconductivity has been achieved. Perhaps the only feature that is generally agreed upon is that high $T_c$ cuprates are not conventional BCS-electron-phonon driven superconductors.

$MgB_2$ offers now the possibility of making real progress in our understanding of superconductivity, by greatly limiting the possibilities for reasonable explanations. The reason is, $MgB_2$ shares many features in common with 'conventional' superconductors, and it also shares many features in common with high $T_c$ cuprates. This leads in our view to just one of the two following possibilities as most compelling:

*(1) $MgB_2$ is a 'conventional superconductor', i.e. electron-phonon driven s-wave, perhaps with some new features due to its high $T_c$, and fundamentally different from the high $T_c$ cuprates.*

*(2) $MgB_2$ is a 'high temperature superconductor', as the high $T_c$ cuprates are, driven by the same physical mechanism.*

We will not consider a third possibility, which in our view is not supported either by experimental evidence or common sense, that a new mechanism different from the conventional electron-phonon one and different from the one operating in high $T_c$ cuprates, is operative in $MgB_2$. While most theorists in the field are favoring possibility *(1)* [2,3,4], the purpose of this paper is to present a theory that favors possibility *(2)* [5], the theory of 'hole superconductivity'[6]. That theory in fact leads to another third possibility, which includes *(2)* but not *(1)* above:

*(3) $MgB_2$, and the high $T_c$ cuprates, and the 'conventional superconductors', <u>and all other superconductors</u>, are driven by the same physical mechanism, which is <u>not</u> the electron-phonon interaction.*

That mechanism, hole undressing, will be reviewed in the following sections. To the extent that superconducting properties of $MgB_2$ resemble those of high $T_c$ cuprates, we argue that possibility *(3)* becomes increasingly likely. However, other theories have also recently proposed that $MgB_2$ and the cuprates are driven by the same (non-conventional) mechanism [7,8].

The following are features of the high $T_c$ cuprates that are shared by $MgB_2$:
- *Planar structure*
- *Transport that drives superconductivity is in-plane*
- *Planes are negatively charged*
- *In-plane transport occurs through overlapping $p_{x,y}$ orbitals*
- *Carriers that drive superconductivity are holes in $p_{x,y}$ orbitals*

If these features are essential to the superconductivity of $MgB_2$ and the cuprates, this would support the proposal *(2)* of a common mechanism for both classes of materials. On the other hand, the following are features of the cuprates not shared by $MgB_2$:
- *Proximity to antiferromagnetism/Mott insulator state*
- *d-electrons at the Fermi surface*
- *Large Hubbard U*
- *Linear resistivity with temperature*
- *Apparent evidence for d-wave symmetry of the superconducting state*
- *Stripes*

If some or all of these features are essential to the superconductivity of the cuprates, as advocated by the majority of other theories proposed to explain their superconductivity (e.g. 'big tent theories'[9]), the mechanism of superconductivity in the cuprates would be fundamentally different from that in $MgB_2$. According to the theory of hole superconductivity, these latter features are in fact <u>irrelevant</u> to the mechanism of superconductivity in the cuprates. Instead, during the past 10 years we have argued that it is the features of the cuprates in the first group, which happen to be <u>precisely</u> the ones that they share with $MgB_2$, that are the essential ones to their superconductivity.

## II. HOLES ARE NOT LIKE ELECTRONS

Central to the understanding of many-electron physics in solids is the concept of a quasiparticle[10]. The quasiparticle is what remains as the coherent part of the bare particle, after taking into account its interaction with other particles. It may be thought of as a particle carrying with it a 'cloud' of other particles with which it interacts. This cloud can be visualized as 'clothing' or 'dressing' of the bare particle, and it will generally lead to an increased effective mass of the quasiparticle compared to the bare particle.

The single particle spectral function for an electron, defined as

$$A(k,\omega) = -\frac{1}{\pi} \operatorname{Im} G(k,\omega) \qquad (1a)$$

$$G(k,\omega) = \int_{-\infty}^{\infty} d\omega e^{i\omega t} < -iTc_{k\sigma}(t)c^+_{k\sigma}(0) > \qquad (1b)$$

(with $T$ the time ordering operator) can be written as

$$A(k,\omega) = z\delta(\omega - \tilde{\varepsilon}_k) + A'(k,\omega) \qquad (2)$$

The first term is the quasiparticle part, giving the fraction $z$ of the bare particle that propagates coherently, and $A'$ is the incoherent contribution due to the interaction of the electron with other electrons. The weight of the incoherent part

$$\int_{-\infty}^{\infty} d\omega A'(k,\omega) = 1-z \qquad (3)$$

will increase as the quasiparticle weight $z$ decreases due to the sum rule Eq. (3). Furthermore, under the assumption that the self-energy is dominantly $\omega$- and not $k$- dependent one has

$$\tilde{\varepsilon}_k = z\varepsilon_k \qquad (4a)$$
$$m^* = m/z \qquad (4b)$$

with $\varepsilon_k$ the bare (non-interacting) electron energy; hence the effective mass renormalization $m^*/m$ is determined by the quasiparticle weight.

Consider now the quasiparticle weight dependence on the electronic occupation of the band, $n$. We denote by $n$ the (average) number of electrons per unit cell in the band, hence $n$ goes from 0 to 2, and $n=1$ denotes a half-filled band. We now argue that the following is generally true for $n<1$:

$$z(n) > z(2-n) \qquad (5)$$

in other words, *z is larger for electrons than for holes*. This is particularly clear in the comparison between a single electron in an empty band ($n->0$) and a single hole in a full band ($n->2$). Whereas for the single electron $z->1$, as it has no other electrons to interact with, for the single hole $z<1$ because a hole cannot be thought of as a completely non-interacting independent particle. A more detailed quantitative argument is given in Ref. [11].

Eq. (5) embodies the fundamental electron-hole asymmetry of condensed matter which is at the heart of the theory of hole superconductivity. When $n<1$ the transport is electron-like, and when $n>1$ it is hole-like. Qualitatively, Eq. (5) implies that electron metals are coherent, since electron carriers are lightly dressed ($z$ large), and hole metals are incoherent since hole carriers are heavily dressed ($z$ small); equivalently, from Eq. (4), that electron carriers are light and hole carriers are heavy. As the filling of a band is increased, electrons become increasingly incoherent as they become increasingly dressed and heavier due to interactions with other carriers in the band, and turn gradually into holes; conversely, as the filling of a band is decreased, holes become increasingly coherent and lighter as they undress and turn into electrons.

Now the essential point that relates this discussion to superconductivity is that pairing of electrons (i.e. when $n<1$) causes the band locally to become more full, i.e. the carriers in the pair to become more hole-like; and conversely, that pairing of holes (i.e. when $n>1$) causes the band locally to become emptier, hence more electron-like. As a consequence, if electrons were to pair they would become more like holes, while if holes pair they become more like electrons. Because electron transport is more coherent than hole transport, the process of hole pairing is favored, as it leads to more coherent transport and lowering of kinetic energy, while the process of electron pairing is disfavored, because it would lead to more incoherent transport and increase of kinetic energy. Hence superconductivity can only occur due to hole pairing, which requires the Fermi level in the normal state to be in the upper half of the band, and in fact, as we will see more quantitatively, close to the top of the band. The qualitative physics resulting from these considerations is shown in Figure 1.

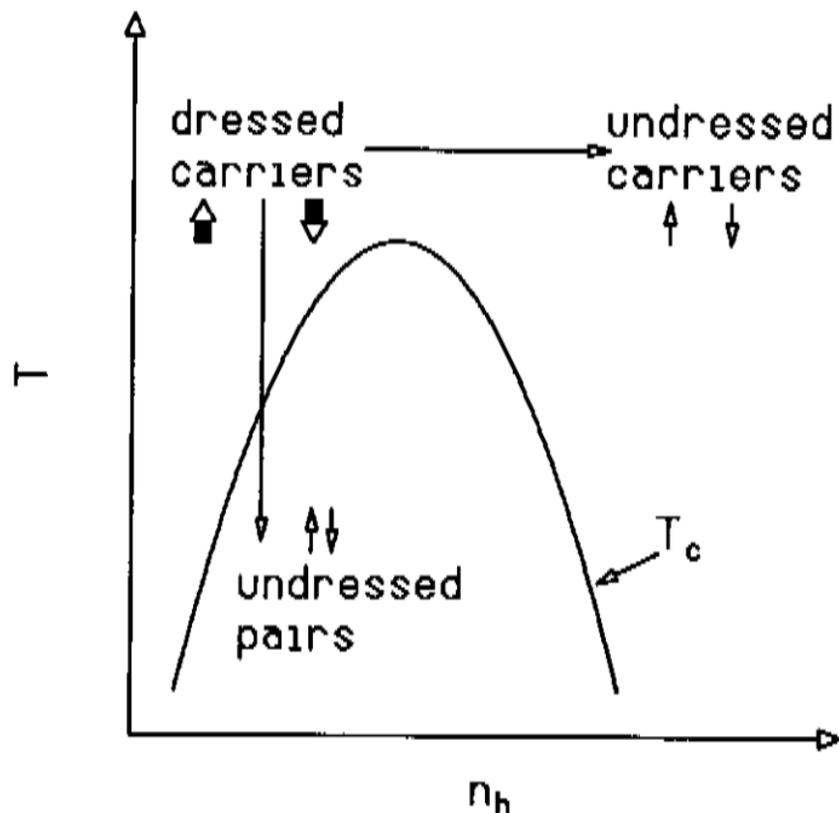

**Figure 1:** Superconducting $T_c$ (schematic) versus hole concentration. $n_h=0$ (origin) corresponds to a full band. Heavily dressed hole carriers at low concentration undress as the temperature is lowered or as the carrier concentration increases.

## III. SUPERCONDUCTIVITY FROM HOLE UNDRESSING

The above considerations are very general and qualitative. We now consider a simple specific realization of these ideas. Assume the operators creating a bare electron and a quasi-electron in the band at site i are related by[12]

$$d^+_{i\sigma} = [1-(1-S)\tilde{n}_{di,-\sigma}]\tilde{d}^+_{i\sigma} \qquad (6)$$

with 0<S<1; the equivalent relation for hole operators is

$$c^+_{i\sigma} = S[1+Y\tilde{n}_{i,-\sigma}]\tilde{c}^+_{i\sigma} \qquad (7a)$$

$$Y \equiv \frac{1}{S} - 1 \qquad (7b)$$

The quasielectron weight is then given by

$$z(n) = [1+(S-1)\frac{n}{2}]^2 \qquad (8a)$$

and is a monotonically decreasing function of $n$, the electron occupation, hence satisfies Eq. (5). Equivalently, the quasihole weight as function of the hole concentration $n_h=2-n$ is given by

$$z(n_h) = S^2[1+Y\frac{n_h}{2}]^2 \equiv z_h[1+Y\frac{n_h}{2}]^2 \qquad (8b)$$

and increases monotonically as the band is doped with holes. The effective mass for holes decreases as the hole concentration increases, and the bandwidth of the coherent quasiholes depends on hole concentration as

$$D(n_h) = D_0 z(n_h) \qquad (9)$$

with $D_0$ the bare electron bandwidth. The qualitative behavior of the band as the hole doping increases is shown in Figure 2. Figure 2 also shows schematically the transfer of spectral weight that occurs from the incoherent part of the spectral function to the quasiparticle (q.p.) peak as $n_h$ increases, as described by the generalized Holstein model discussed in Ref. 12.

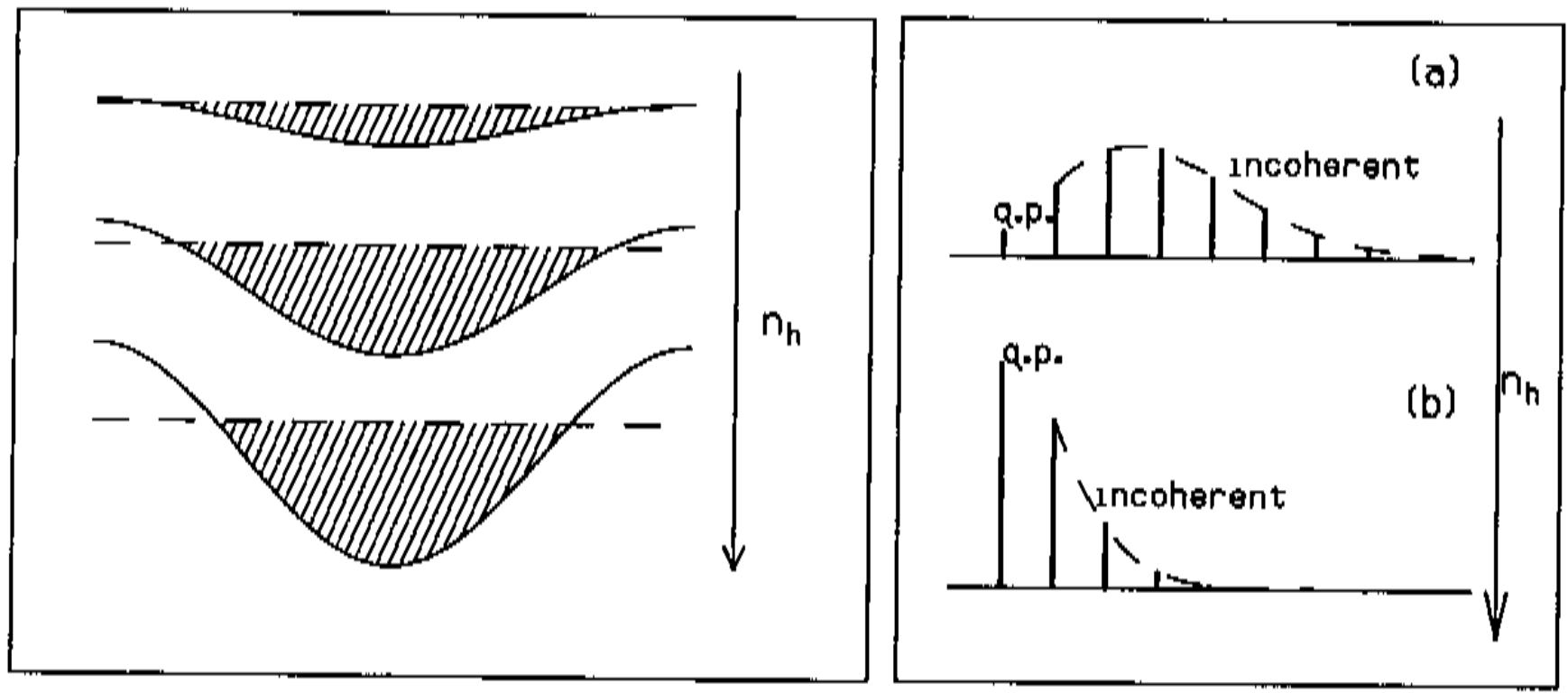

**Figure 2**: Electronic energy band (schematic) (left). As the system is doped with holes, the Fermi level decreases and the bandwidth increases. On the right, the transfer of spectral weight in the single particle spectral function is shown[12].

The 'undressing parameter' Y (Upsilon) (Eq. (7b)) indicates both how heavily dressed the holes in the nearly filled band are, and how rapidly they undress as the hole concentration increases. It is the fundamental parameter in the theory, that determines whether the system is in a strong coupling regime (large Y) or in a weak coupling coupling regime (small Y). In particular, large (small) Y gives rise to high (low) $T_c$ and small (large) coherence length in the superconducting state.

If we now consider the kinetic energy of holes in a tight binding formulation

$$H_{kin} = -\sum_{ij\sigma} t_{ij}^0 (c_{i\sigma}^+ c_{j\sigma} + h.c) \qquad (10)$$

and express the bare hole operators in terms of the quasihole operators Eq. (7) we obtain

$$H_{kin} = -\sum_{ij\sigma} t_{ij}^\sigma (\tilde{c}_{i\sigma}^+ \tilde{c}_{j\sigma} + h.c) \qquad (11a)$$

$$t_{ij}^\sigma = t_{ij}[1 + Y(\tilde{n}_{i,-\sigma} + \tilde{n}_{j,-\sigma}) + Y^2 \tilde{n}_{i,-\sigma} \tilde{n}_{j,-\sigma}] \qquad (11b)$$

$$t_{ij} = t_{ij}^0 S^2 \qquad (11c)$$

The 'correlated hopping' parameter

$$\Delta t_{ij} = Y t_{ij} \qquad (12)$$

gives the increase in hopping amplitude of paired holes compared to single holes, and gives rise to superconductivity even in the presence of substantial Coulomb repulsion[13].

The relations Eq. (6) or (7) arise from detailed atomic physics considerations as discussed in refs.[14-17]. In particular, for the simple case of hydrogen-like ions with nuclear charge Z we obtain[11]

$$S = \frac{(1 - \frac{5}{16Z})^{3/2}}{(1 - \frac{5}{32Z})^3} \qquad (13)$$

This implies that when the effective nuclear charge Z is small, S becomes small and hence Y becomes large, giving rise to high temperature superconductivity. This situation corresponds to the ion being negatively charged, and is due to the fact that the orbitals become more 'floppy' in that case. The relations (8b) and (4) then imply that high temperature superconductivity will generally be associated with incoherent transport in the normal state.

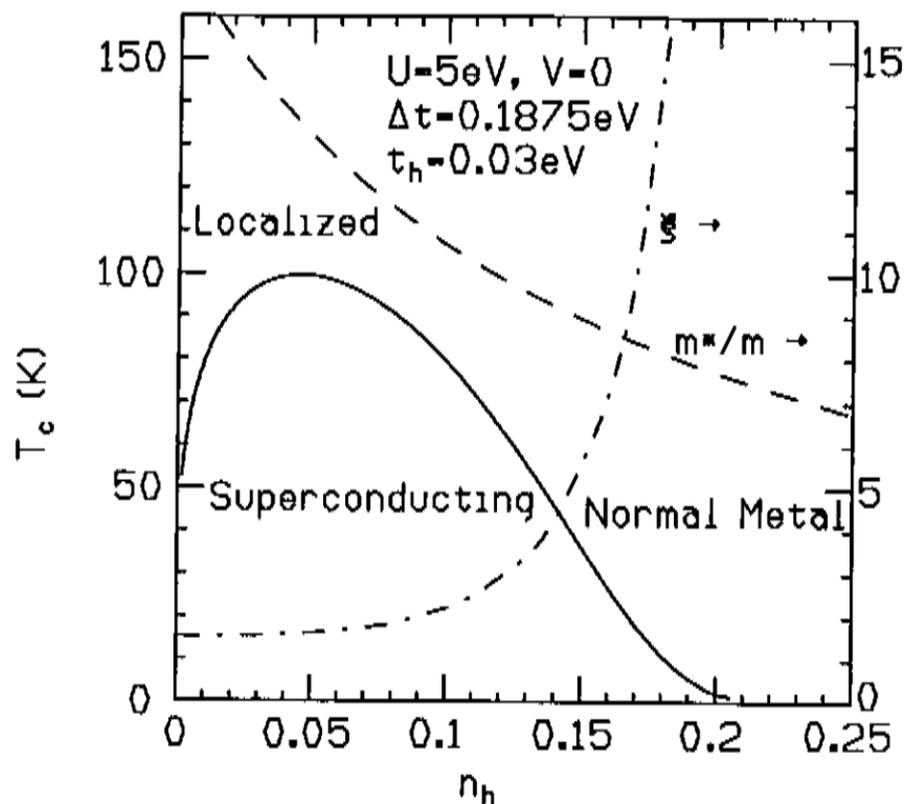

**Figure 3:** typical behavior of $T_c$ versus hole concentration obtained from the model Eq. (11), with added on-site and nearest neighbor Coulomb repulsions U and V. The behavior of the effective mass versus doping and the superconducting coherence length versus doping are also shown.

The typical behavior of $T_c$ versus hole concentration resulting from this physics is shown in Figure 3, for a square lattice with nearest neighbor hopping, with $t_{ij}=t_h$, $\Delta t_{ij}=\Delta t$. $T_c$ increases first as holes are added to a full band, then decreases for higher hole doping, as the undressing becomes less effective and can no longer overcome the cost in Coulomb potential energy that occurs upon pairing. The normal state effective mass decreases monotonically with hole doping, and the superconducting coherence length increases with hole doping, as a cross-over from strong to weak coupling regimes takes place[13].

### IV. HOLE SUPERCONDUCTIVITY IN THE CUPRATES

The theory outlined above was developed and applied to the cuprates long before $MgB_2$ was discovered[13-17]. We assumed at the outset[18] that the relevant carriers in the CuO planes that drive superconductivity are holes doped into planar oxygen $p\pi$ orbitals, which hop through direct orbital overlap from an $O^=$ ion to a nearest neighbor $O^=$ ion, as shown in Figure 4 (left) . These $p\pi$ orbitals are orthogonal to the oxygen $p\sigma$ orbitals that are hybridized with the Cu $d\_x^2-y^2$ orbitals that are believed to be the relevant ones in essentially all other theories of high $T_c$ superconductivity. The presence of antiferromagnetism nearby in the phase diagram, arising from these latter orbitals, is essentially irrelevant within our theory. Hole doping of the oxygen $p\pi$ orbitals brings the Fermi level down from the top of a full band (hereafter called O band) , and at the same time presumably closes the Mott-Hubbard-Slater antiferromagnetic gap in the $d\_x^2-y^2$ band (herafter called Cu-O band), as the entire system becomes a Fermi liquid. This is show schematically on the right side in Figure 4.

While superconductivity is determined by the O $p\pi$ hole carriers and can be understood with a single band model[13], we have also studied the properties of a two-band model as depicted in Fig. 4[19]. Superconductivity in the Cu-O band (electron band) is driven py pairing of holes in the O band (hole band); while $T_c$ doesn't change much due to interband coupling, a second smaller gap is found to appear due to the weaker induced pairing in the electron band. Such a situation could also arise in $MgB_2$, as discussed in the next section.

The experimental observations in the cuprates are complicated by the fact that there are two types of carriers at the Fermi energy, and we do not claim to understand them all. This is augmented by the fact that there should be strong antiferromagnetic fluctuations for the carriers in the Cu-O band due to the proximity of the antiferromagnetically ordered state. Nevertheless, we expect the qualitative physics of the hole carriers in the O band, described in the previous section, to be evident and determine the properties of the cuprates that are relevant to superconductivity. This is indeed the case, as summarized briefly below.

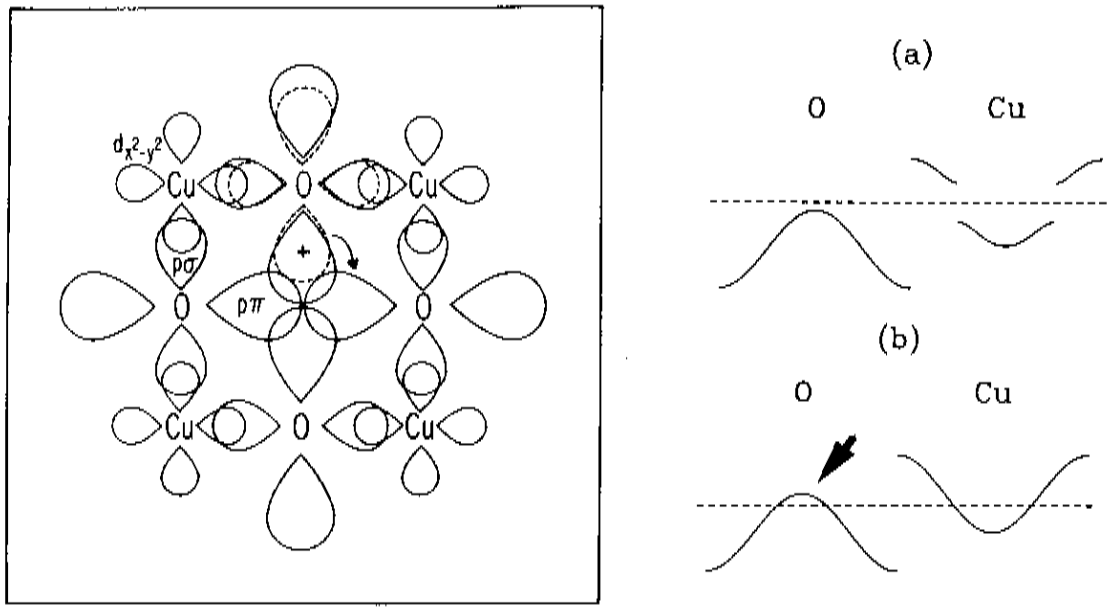

**Figure 4: (left)** Schematic representation of the Cu d orbitals and the O pσ and pπ orbitals. The holes responsible for superconductivity occupy the pπ orbitals and hop through their direct overlap[18]. The dashed and full lines indicate the contraction and expansion of the orbitals that occurs as the holes hop and give rise to the hole dressing. **(right)**: Schematic depiction of the O pπ band (labeled O) and the Cu d-Opσ band (labeled Cu). The former is full, the latter is half-filled with a presumed Mott-Hubbard gap, in the undoped system. Upon doping with holes, the Fermi level shifts down into the O pπ band (and the band expands, not shown here); this is the band responsible for superconductivity, as indicated by the arrow. Also, the gap in the Cu band disappears.

**1) Hall coefficient in the normal state is positive:** this is direct evidence that the carriers are dominantly hole-like, namely the oxygen hole carriers in the O band discussed above. Even in electron-doped cuprates, it has been found in a variety of transport experiments that superconductivity only appears when hole carriers start to dominate the transport[20]. We have predicted this effect[21], and discussed in detail how induced hole carriers will appear in the T' crystal structure of the electron-doped oxides when they are doped with electrons[22].

**2) The normal state is very incoherent for low hole doping:** this is evident from a variety of experimental studies, e.g. the large normal state resistivity, the small low energy (Drude) optical absorption, corresponding to intra-band processes, and the large optical absorption at larger energies, and the broad incoherent photoemission spectra. It is evidence for the small quasiparticle weight of the holes, $z_h$, which will be the case if $S \ll 1$, or equivalently $Y \gg 1$. If the quasiparticle weight is sufficiently small it will not be visible and the system will appear to be completely incoherent and non-Fermi-liquid like. According to Eq. (13), small $z_h$ occurs when the effective nuclear charge of the ions is small, i.e. negatively charged ion. This is precisely the situation in the Cu-O planes where the relevant hole carriers occur in the highly negatively charged $O^=$ ions.

**3) The normal state becomes increasingly more coherent as holes are added:** as predicted by Eq. (8), 'undressing' should occur in the normal state as the hole concentration increases: the system should become more coherent as holes are added and the quasiparticle weight increases, and the quasiparticle effective mass should decrease. This physics is seen in a variety of observations, namely: (i) The resistivity decreases more rapidly than just due to the increased number of carriers[23], indicating that the effective mass of the carriers is decreasing; (ii) Spectral weight in optical absorption in the normal state is transferred from high frequencies to low frequencies as the system is doped with holes[24]; (iii) In photoemission, no quasiparticle peak is seen in the underdoped regime in the normal state and a quasiparticle peak arises in the overdoped regime.[25]

**4) The system becomes more coherent as it goes superconducting:** it has been repeatedly observed that high $T_c$ cuprates are 'strange' in the normal state, especially if underdoped, and become less strange, i.e. more Fermi liquid like, as they go superconducting. This is precisely what the theory discussed here predicts. The quasiparticle weight increases upon pairing because the 'local' hole concentration increases; it is as if in Eq. (8b), $n_h$ is changed to $n_h+1$ upon pairing. Experimentally, this is seen in photoemission as a sharp quasiparticle peak emerging from the incoherent background as the temperature is lowered and the system becomes superconducting[26]; calculation of the spectral function within the theory discussed here shows a similar effect[12].

**5) Coherence in the superconducting state increases with hole doping:** it is seen in photoemission that the quasiparticle weight in the superconducting state (the spectral weight in the sharp peak in the photoemission spectrum near zero energy) increases linearly with hole doping, and levels off in the overdoped regime[26,27]. Again, the increase with hole doping $n_h$ is qualitatively expected from Eq. (8), and the leveling off for high doping is also reproduced by the theory[12].

**6) The carrier effective mass becomes smaller as the system goes superconducting:** this is directly related to the increased quasiparticle weight discussed in **4)**, as given by the relation (4b). It can equivalently be understood from the increase in hopping amplitude from $t_{ij}$ to $t_{ij}+\Delta t_{ij}$ that occurs upon pairing. We have pointed out that this should lead to an apparent violation of the Ferrell-Glover-Tinkham optical sum rule, as spectral weight in optical absorption should be transferred from high frequencies to the zero frequency δ-function as the system goes superconducting[28]. Both the 'color change' resulting from decrease of high frequency absorption [16] as well as the violation of the low energy optical sum rule[28] predicted have been seen experimentally[29,30,31].

**7) Positive pressure dependence of $T_c$:** the parameter that drives superconductivity, $\Delta t$, increases as the distance between O atoms in the plane decreases, because it depends on wavefunction overlaps as the single particle hopping $t$ does. Hence, $T_c$ is found to increase strongly with pressure applied in the planes in this theory[13], while the theory predicts that pressure in direction perpendicular to the planes has a much smaller effect and can lead to either increase or decrease of $T_c$ depending on parameters. In the high $T_c$ cuprates, extensive experiments both with isotropic pressure and uniaxial pressure have been performed[32]. Pressure in the planes is found to always increase strongly the transition temperature, in accordance with the theory. In interpreting pressure experiments care needs to be used to disentangle the effect of possible charge transfer to and from planes with pressure[33]. It is found in fact that 1GPa of pressure can lead to about a 10% increase in hole concentration in the planes. This will lead to either increase or decrease of $T_c$ depending on whether the system is in the underdoped or overdoped regime, and needs to be taken into account to extract the intrinsic pressure dependence of $T_c$.

**8) Charge asymmetry effects:** the theory of hole superconductivity is based on the fundamental asymmetry between electrons and holes, which should have direct observable consequences in charge asymmetry properties. One prediction of the theory is asymmetric NIS tunneling characteristics, with larger conductance for a <u>negatively</u> biased sample[34]. This prediction seems to be increasingly confirmed by tunneling experiments. The theory also predicts an intrinsic positive thermoelectric power for NIS tunnel junctions[35], a prediction that has not yet been experimentally tested. Furthermore, the theory predicts that negative charge should be expelled from the bulk of the superconductor towards the surface[36]; in the mixed state, this should lead to negatively charged vortices and positively charged superconducting regions. There is some evidence in support of this, although other evidence appears to contradict it[37].

**9) Charge inhomogeneity:** the theory predicts a large sensitivity to non-magnetic disorder, that can lead to inhomogeneous charge distribution and local gap variations[38]. Some evidence for this is seen in recent STM tunneling experiments[39].

In conclusion, the theory of hole superconductivity predicts a remarkable number of qualitative features that resemble qualitative features of the high $T_c$ cuprates. Furthermore, quantitative calculations yield results that agree with experiment, using some experimental observation(s) to fit the parameters in the theory that are difficult to predict accurately from first principles. The parameters thus obtained are in the range of plausible parameters predicted by microscopic calculations. It should also be pointed out that the predictions that emerge from the theory are inescapable, determined from the single assumption Eq. (5), as implemented by Eq. (7). Furthermore, this assumption is supported by direct microscopic analysis of the physics of electrons in atoms starting from first principles[17].

The theory predicts unequivocally that the symmetry of the superconducting state is s-wave, which appears to be contradicted by much experimental evidence, but is also supported by some experimental evidence[40].

**V. HOLE SUPERCONDUCTIVITY IN MAGNESIUM DIBORIDE**

The discovery of a single material where superconductivity occurs in the absence of hole carriers at the Fermi energy would prove the theory outlined above wrong. In particular, when $T_c$ is high the dominance of hole carrier transport should be particularly evident. It is then not surprising that this is the case in $MgB_2$[41], which has a $T_c$ substantially higher than other similar intermetallic compounds.

To understand the physics of superconductivity in MgB$_2$ we take a Cu-O plane and strip it of irrelevant elements, namely we discard the Cu. We then change O$^=$ into B$^-$, and change the structure from square to honeycomb. The different structure, with two atoms in the unit cell, leads to two, slightly different, hole bands arising from overlaping planar $p$ orbitals rather than just one hole band as in the cuprates, but is not an essential difference. If these hole bands are almost full, as the O $p\pi$ band discussed in the previous section, superconductivity will arise in the B$^-$ planes driven by pairing of hole carriers in the $p_{xy}$ orbitals. Because the B$^-$ ions are negatively charged but less so than the O$^=$ ions, superconductivity should occur at high temperatures but not as high as in the cuprates.

Figure 5 shows an example of band structure calculation results for MgB$_2$[42]. The hole pockets at the Γ point in two of the bands are clearly visible. These two bands give rise to almost cylindrical sheets of the Fermi surface, indicating that transport of these hole carriers is essentially two-dimensional. It is remarkable that there is general agreement, by workers that hold completely different views on the origin of superconductivity in MgB$_2$, that the relevant carriers that drive superconductivity are indeed these holes in B$^-$ $p_{xy}$ orbitals, that hop through direct orbital overlap from a B$^-$ ion to a neighboring B$^-$ ion, just as we assumed for the high T$_c$ cuprates. How are we then to tell the different theories apart? We discuss here several criteria:

1) **Dependence of T$_c$ on doping:** The theory of hole superconductivity predicts that T$_c$ should follow the 'universal' bell-shaped curve observed for T$_c$ versus doping in the cuprates and in the transition metal series (discussed in the next section). From band structure calculations we infer that MgB$_2$ is slightly overdoped, as shown in Figure 6, with average hole content per boron atom n$_h$=0.065. Electron-phonon calculations have not provided similar explicit predictions of T$_c$ versus doping. However, from the published results for the density of states and assuming that other parameters in electron-phonon theory (electron-phonon coupling, phonon frequencies and Coulomb pseudopotential) do not vary strongly with doping we infer the dependence of T$_c$ with doping within electron-phonon theory should be approximately as shown in Fig 6 [5]. It can be seen that the predictions of electron-phonon and the present theory are qualitatively different.

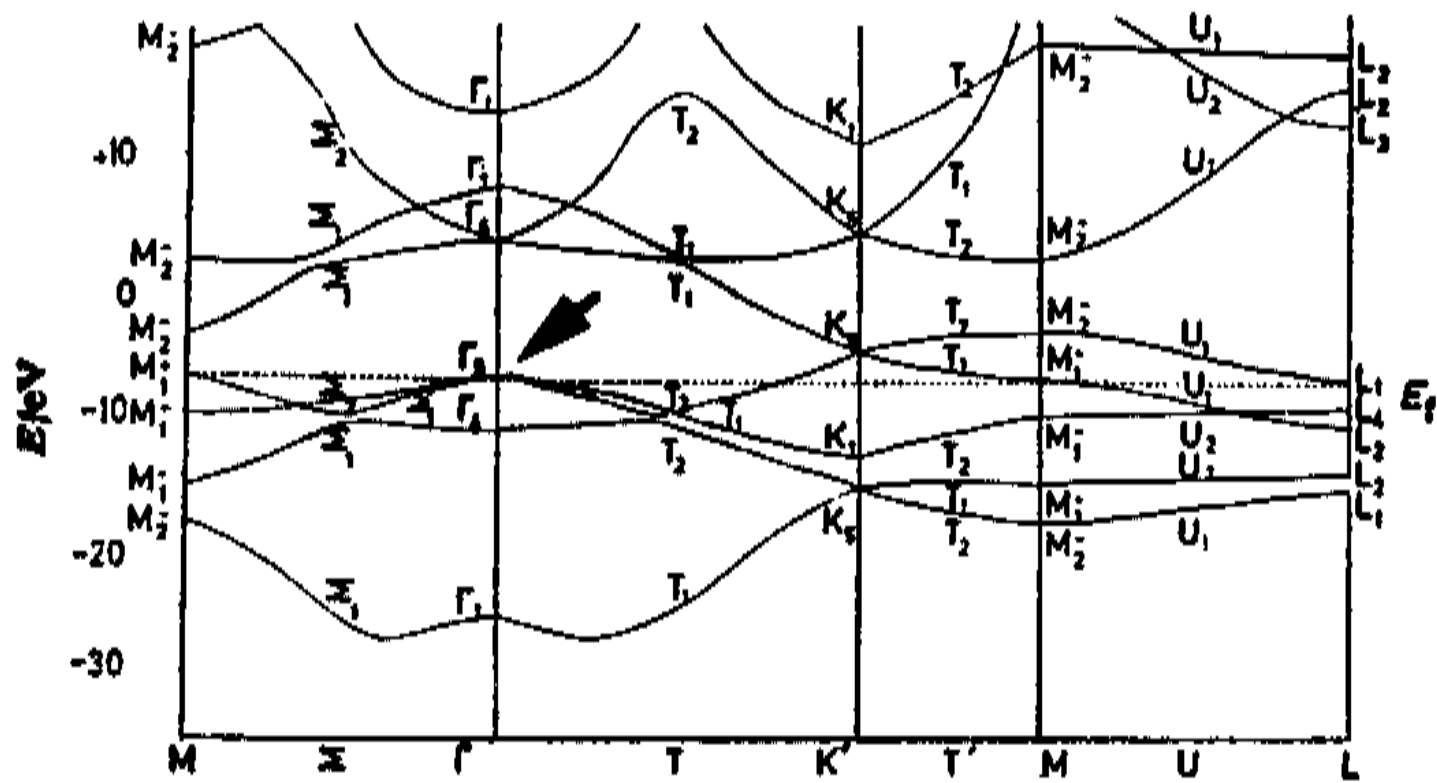

**Figure 5:** Results from band structure calculations for MgB$_2$ in the plane directions, reproduced from Ref. [42] by Armstrong et al. Note the hole pockets at the $\Gamma_5$ point (where the arrow points), that give rise to cylindrical hole-like Fermi surfaces.

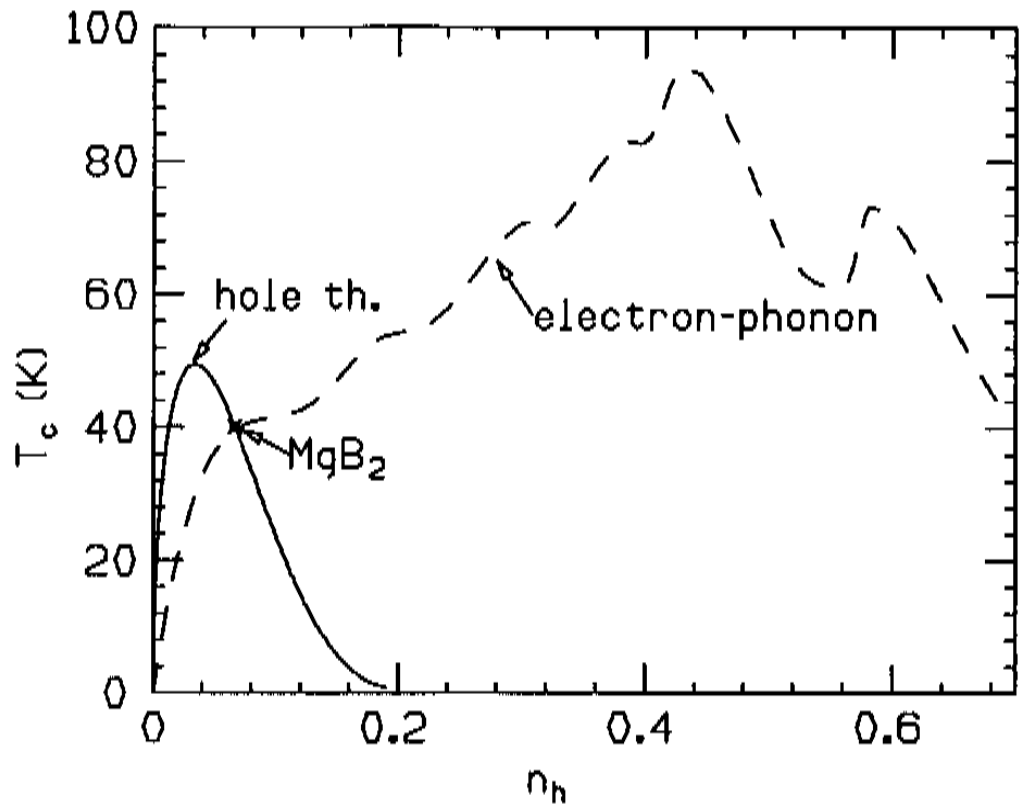

**Figure 6:** Prediction for the variation of the critical temperature with hole doping in the model of hole superconductivity (full line). The expected behavior from electron-phonon theory, assuming constant electron-phonon matrix elements, phonon frequencies and Coulomb pseudopotential is shown as the dashed line.

In particular,

$$\frac{\partial T_c}{\partial n_h}\bigg)_{MgB_2} < 0 \qquad \text{hole theory} \qquad (14a)$$

$$\frac{\partial T_c}{\partial n_h}\bigg)_{MgB_2} > 0 \qquad \text{electron-phonon theory} \qquad (14b)$$

so that a small increase in the hole concentration of $MgB_2$ should lead to a decrease of $T_c$ within the hole theory, and to an increase in $T_c$ within electron-phonon theory. Furthermore, the hole theory predicts that $T_c$ will drop rapidly to zero as holes are added, while electron-phonon theory predicts that $T_c$ should remain large for a much larger range of hole doping. Substituting Mg by an alkali metal like Li should lead to an increase of 1 hole per unit cell; approximately 1/6 of this should be the increase in $n_h$ per B atom, which will drive $T_c$ to zero within our theory. On the other hand, it would appear from Fig. 6 that electron-phonon theory would predict a substantially higher $T_c$ for $LiB_2$ than for $MgB_2$. To our knowledge, no explicit prediction of $T_c$ for $LiB_2$ within electron-phonon theory has been published. A recent experimental paper reports $T_c=0$ for $LiB_2$[43].

**2) Dependence of $T_c$ on pressure:** Experiments report decreasing $T_c$ under application of hydrostatic pressure, which appears to be in agreement with the prediction of electron-phonon theory[44]. The hole theory predicts that $T_c$ should increase if the distance between B atoms in the plane decreases. Unfortunately, uniaxial pressure experiments have not yet been performed. Under hydrostatic pressure it is possible that the reduction in the c-axis distance is a dominant effect, and also that significant charge transfer to or from the planes occurs. Hence the observed suppression of $T_c$ under hydrostatic pressure is not necessarily in contradiction with our theory.

**3) Dependence of other observables on doping:** we have recently discussed the expected behavior of a variety of observables with doping for compounds derived from $MgB_2$[5]. In particular, coherence length should increase monotonically with hole doping and penetration depth should decrease monotonically with hole doping (assuming the system remains in the clean limit). Experiments that would test these predictions have not yet been reported. Similar predictions have not been made within electron-phonon theory.

**4) Tunneling asymmetry:** our theory predicts asymmetry in tunneling of the same sign but smaller magnitude than in the cuprates. The experimental situation is unclear.

**5) Isotope effect:** a B isotope coefficient α~0.26 has been reported experimentally[45]. Both electron-phonon theory and the hole theory predict a positive isotope effect. Within the hole theory it is difficult to calculate α but simple estimates suggest it could have a range of values and even be larger than 0.5[5]. Within electron-phonon theory it is not easy to explain an α as small as observed, since it would require a large value of the Coulomb pseudopotential, that would in turn lead to a smaller $T_c$. Furthermore, for the compound $BeB_2$ an isotope effect α~1 has been reported[46], and a $T_c$ of only ~ 1K. This is even more difficult to explain within electron-phonon theory.

**6) Symmetry of the superconducting state:** both the electron-phonon theory and the hole theory predict s-wave superconductivity. Most experimental results so far appear to indicate an s-wave state. There are however some indications of anomalous behavior similar to observations in the cuprates, which in the cuprates has been interpreted as evidence for nodes in the gap, for example in penetration depth measurements[47].

**7) Multi-band and anisotropy effects:** both the hole theory and electron-phonon theory could give rise to similar effects due to the presence of multiple bands at the Fermi energy. The hole theory predicts that the pairing potential in the $p_{xy}$ hole bands should be much larger than in the $p\pi$ bands ($p_z$ orbitals), where the carriers are electron-like[19]. Similarly, electron-phonon theory appears to predict stronger electron-phonon coupling for $p_{xy}$ holes than for carriers in the pπ bands. Furthermore, the $p_{xy}$ bands are two-dimensional and the $p\pi$ bands are three-dimensional. Hence both theories could give rise to similar effects, for example different size gaps arising from carriers in the different bands, and anisotropic behavior in properties associated with superconductivity. Anisotropy in superconducting properties such as $H_{c2}$ has recently been reported[48]. The experimental situation concerning multiple gaps is unclear.

In summary, it appears that future experiments in $MgB_2$, particularly as function of doping, should be able to shed light onto whether the electron-phonon theory or the hole theory are more appropriate to describe superconductivity in $MgB_2$ and related compounds. It would be very interesting to electric-field dope $MgB_2$[49], to test the relation Eq. (14) without additional effects such as disorder that could arise with chemical doping.

Thus, according to the theory of hole superconductivity, $MgB_2$ is a pristine realization of the essential physics that drives superconductivity in the high $T_c$ cuprates. It has conducting substructures (planes) formed by tightly packed negative ions ($B^-$), where conduction occurs through holes in nearly filled bands, that arise from direct overlap of orbitals on neighboring negative ions. The physics of superconductivity is the same as in the high $T_c$ cuprates, without the complications introduced by the Cu ions, which we have argued are irrelevant to the physics of superconductivity in the cuprates[11-19]. The difference between superconductivity in $MgB_2$ and the cuprates is only that the 'undressing parameter' Y that drives superconductivity is smaller in $MgB_2$, leading to a weaker coupling regime and larger coherence length.

## VI. HOLE SUPERCONDUCTIVITY IN OTHER MATERIALS

A compelling aspect of the theory discussed here is that it is proposed to be a *universal* theory of superconductivity, applicable to all materials. This is so because it is based on very general principles, as outlined in previous sections. Thus, the theory would appear to be easily falsifiable, by finding a single material that does not have hole carriers at the Fermi energy. Unfortunately because Fermi surfaces are usually complex and have both electron- and hole-like portions, this is not a trivial task.

Nevertheless, we argue that examination of the systematics of superconductivity in many materials, and in particular in the elements, provides strong empirical support for our theory. Consider for example the band structures of two elements shown in Figure 7 below[50]. Several of the characteristics of these two elements suggest within electron-phonon theory that element 1 should have the highest transition temperature, namely[51]:
- Ratio of densities of states at the Fermi level (element 1 / element 2) =**1.24** (1.73 states/eV-atom / 1.40 states/eV-atom).
- Ratio of Debye temperatures ( 1 / 2 ) = **1.62** (450K / 277K).
- Ratio of atomic masses ( 1 / 2) = **0.48.**
- Ratio of temperature-dependence of resistivity (1 / 2)=**2.6** (1.1 / 0.42).

(The high temperature dependence of the resistivity indicates the strength of the electron-phonon coupling[51].)

On the other hand, a glance at the band structures in Fig. 7 indicates instead that within the hole theory, element 2 should be superconducting, as it shows a hole pocket at the Γ point (as $MgB_2$ does), and element 1 should be non-superconducting as it shows no hole-like states at the Fermi energy.

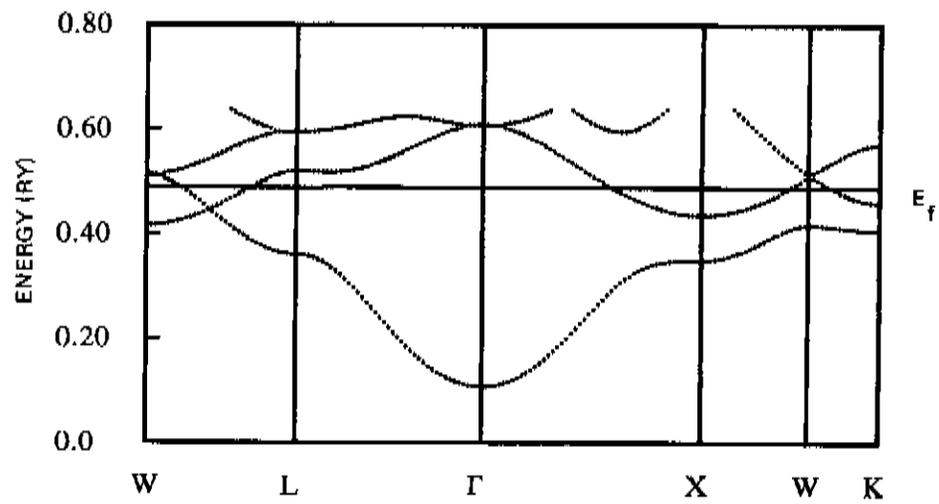

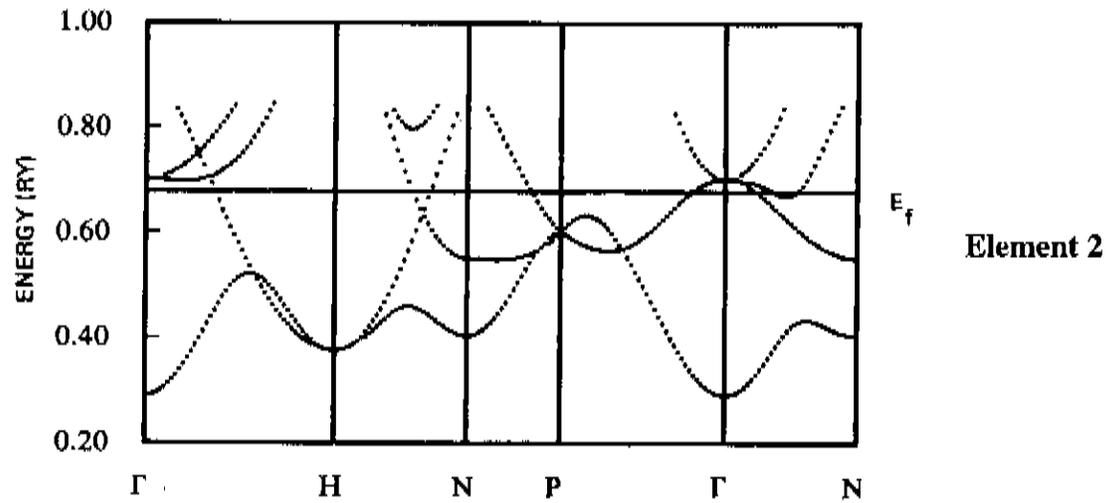

Figure 7: band structures of two elements (from ref. [50]).

Indeed, the measured Hall coefficient is positive for element 2 and negative for element 1. And indeed, element 2 is Nb, with $T_c$=9.5K the highest $T_c$ among the elements (under atmospheric pressure) and element 1 is Sc, a non-superconductor. It has been pointed out by Gladstone, Jensen and Schrieffer [51] that the absence of superconductivity in Sc is somewhat of a puzzle within electron-phonon theory .

The correlation between the sign of the Hall coefficient and occurrence of superconductivity in elements and compounds is very strong, as pointed out by Chapnik and others[52,53]. For the elements, Figure 8 below shows $T_c$ versus Hall coefficient $R_H$. Out of 44 non-magnetic elements, 25 are superconductors and 19 are non-superconductors. Out of the 25 superconductors, 18 have $R_H >0$, 5 have $R_H<0$ (Ga, Sn, La, Hf, Hg); we were unable to find data for $R_H$ for Tc and Os. Out of the 19 non-superconductors, 16 have $R_H<0$, 1 has $R_H>0$, and we were unable to find data for $R_H$ for Sr and Ba. Thus, the correlation between positive Hall coefficient and superconductivity in the elements is very strong, the probability that this correlation is due to chance is less than 0.00001[54]. So far no other theory has proposed an explanation for this correlation. The theory discussed here predicts that for simple band structures superconductors should have positive Hall coefficient and non-superconductors should have negative Hall coefficient. For complicated band structures it is however possible that superconductivity is driven by the hole-like parts of the Fermi surface, yet normal state transport is dominated by electron-like carriers at the Fermi energy resulting in a negative Hall coefficient.

The behavior of $T_c$ versus electron concentration in transition metal alloys also provides strong support for the theory discussed here. There are two bell-shaped curves as the number of electrons per atom $n_e$ increases, at $n_e$~4.7 and $n_e$~6.5 , as seen in Figure 9 (Matthias' rules)[55]. The behavior is similar to the $T_c$ versus hole concentration dependence seen in the high $T_c$ oxides. While some correspondence with the behavior of the density of states has been argued[56], density of states can certainly not explain why $T_c$ is not high both at the beginning and the end of the transition metal series, where the density of states is large[51]. Within the theory of hole superconductivity, the two bumps arise as the Fermi level approaches the top of the d $t_{2g}$ bands at the $\Gamma$ point ($\Gamma_{25'}$) and the $e_g$ bands at the $\Gamma$ point ($\Gamma_{12}$). $T_c$ goes to zero when there is no band that is almost full for given position of the Fermi level[21,57].

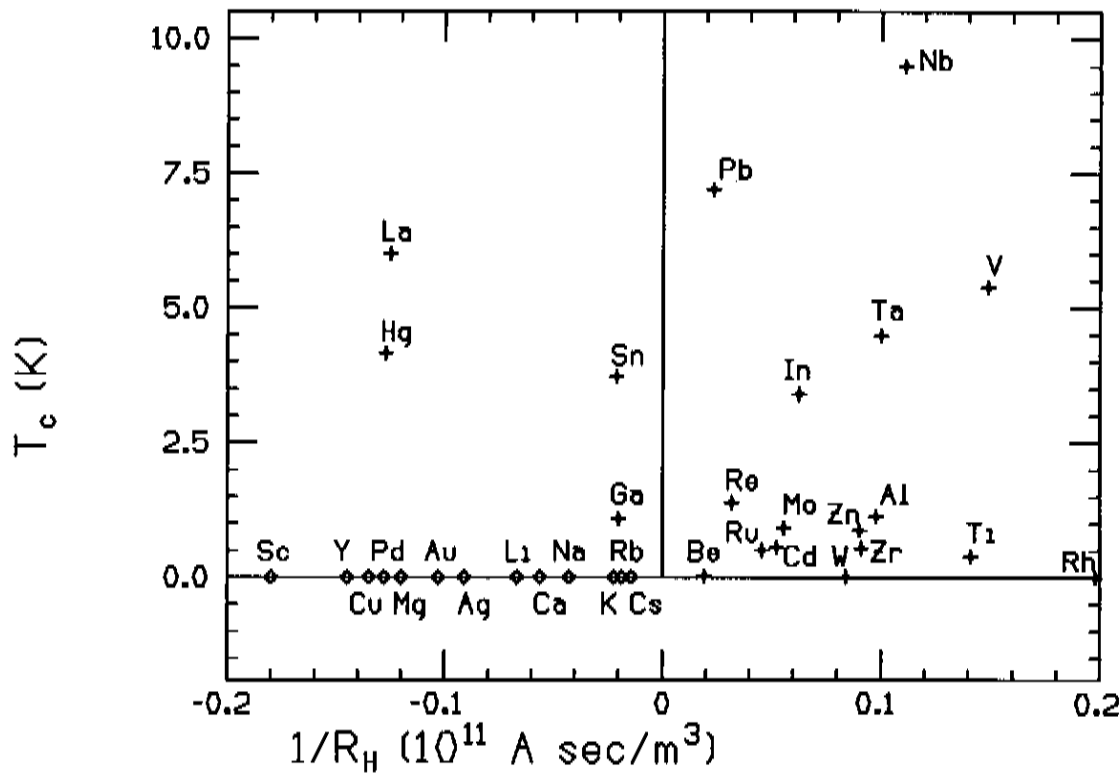

**Figure 8:** $T_c$ versus inverse Hall coefficient for the non-magnetic elements. 7 elements are missing from the graph because $R_H$ is outside the scale: As, Pt, Bi, non-superconductors with $R_H<0$; Ir and Tl, superconductors with $R_H>0$; Sb, non-superconductor with $R_H>0$, and Hf, superconductor with $R_H<0$. For Sr, Tc, Ba and Os we were unable to find data for $R_H$.

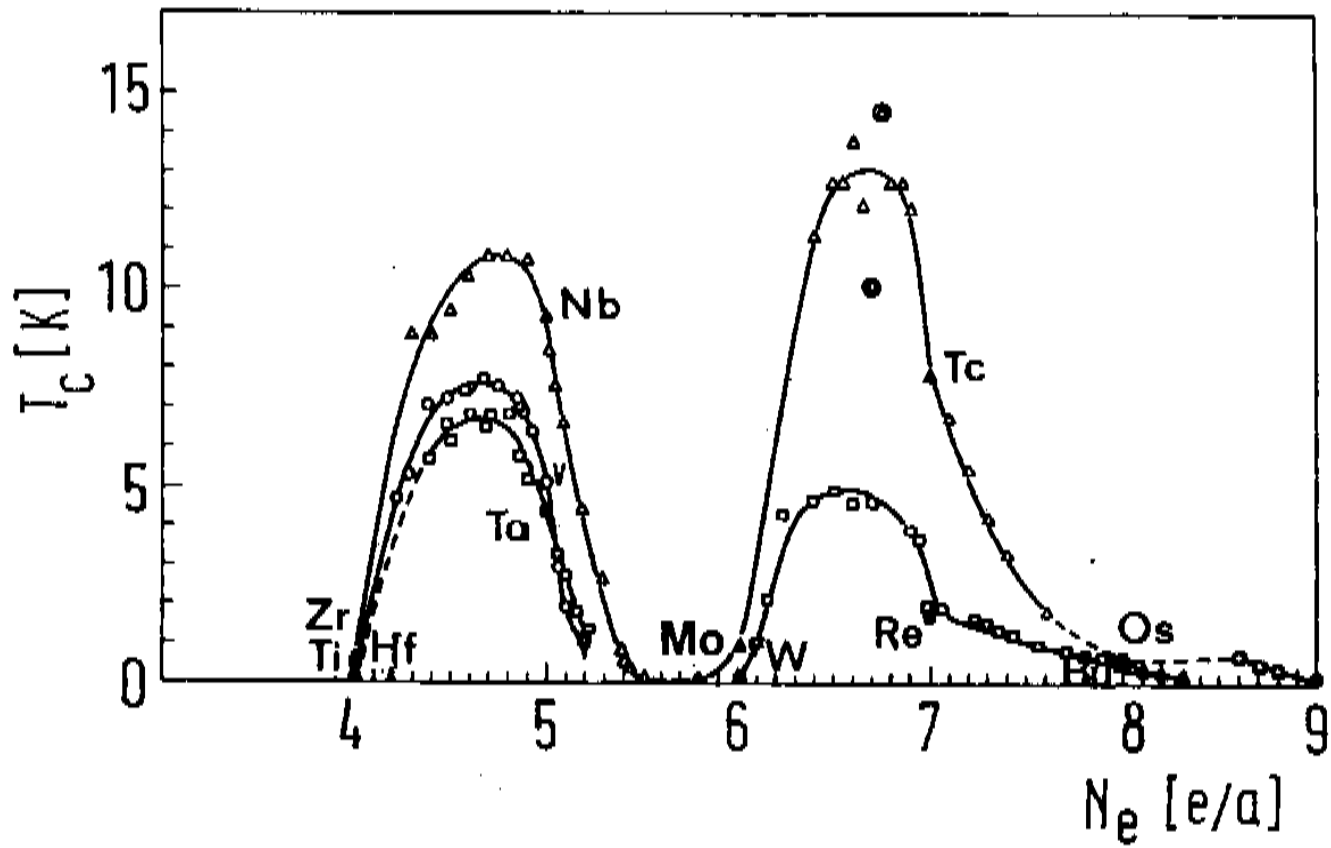

**Figure 9:** The dependence of $T_c$ on the number of valence electrons, $N_e$, in solid solutions formed by neighboring transition elements, from Ref. 55.

In Figure 10 below we show in more detail the dependence of $T_c$ on electron number for the first bump for elements in the 4th row of the periodic table, together with the band structures for Ti, V and Cr around the Γ points.

There are many other anomalies in the occurrence of superconductivity that are not easily explained by electron-phonon theory. Lithium has been long predicted to be a superconductor[59], yet is found to be non-superconducting to ever lower temperatures. In our theory this is naturally explained by the fact that Li does not have holes at the Fermi energy. Beryllium has a very low $T_c$ in the bulk, and $T_c$ increases by a factor of 1000 in thin films. In our view, Be is a semimetal with very small hole pockets, and a small shift in the Fermi level when it is in thin film form can substantially increase the number of hole carriers at the Fermi energy. Indeed, measurement of the Hall coefficient (positive) indicate a large change in magnitude between bulk and thin film forms[60]. We believe it should be possible to form Be compounds that optimize the number of holes at the top of the 2s band and have substantially higher $T_c$ than bulk Be.

Finally, recent experiments in electric-field doped $C_{60}$[61] provide striking evidence for the importance of electron-hole asymmetry discussed in this paper. Both hole-doped and electron-doped $C_{60}$ show a bell-shaped $T_c$ versus doping that resembles the one seen in high $T_c$ cuprates and in the transition metal series, however for the hole-doped case $T_c$ is 5 times larger. Furthermore, the resistivity in the normal state is substantially larger for the hole-doped case, which we interpret to be due to enhanced dressing of hole versus electron-carriers as discussed earlier in this paper. Even for the electron-doped case we believe that superconductivity is due to some minority hole carriers that are induced upon electron-doping, similarly to what happens in the electron-doped cuprates.

## VII. CONCLUSIONS

The theory of hole superconductivity is based on the fundamental electron-hole asymmetry of condensed matter, that has its origin in the fact that the positive proton is two thousand times heavier than the negative electron. Because it is rooted in fundamental principles, it cannot possibly apply to some materials and not to others: it either applies to all materials, or to none. Therein lies its greatest strength, but also its weakness, since it can be temporarily dismissed by pointing at examples that seemingly contradict it. For example, electron-doped cuprates appeared initially to contradict the theory[62], until more careful experiments were performed[20].

Fortunately, $MgB_2$, because of its simplicity, provides an ideal testing ground, and is immediately seen to fit the theory very well[5]. For all other theories of superconductivity, the finding that in $MgB_2$ high $T_c$ superconductivity results from hole carriers in nearly full bands propagating through negatively charged ions is incidental, not sufficient and certainly not necessary for high $T_c$. For the theory of hole superconductivity instead it is a prediction, even if it was not explicitly made for this particular compound.

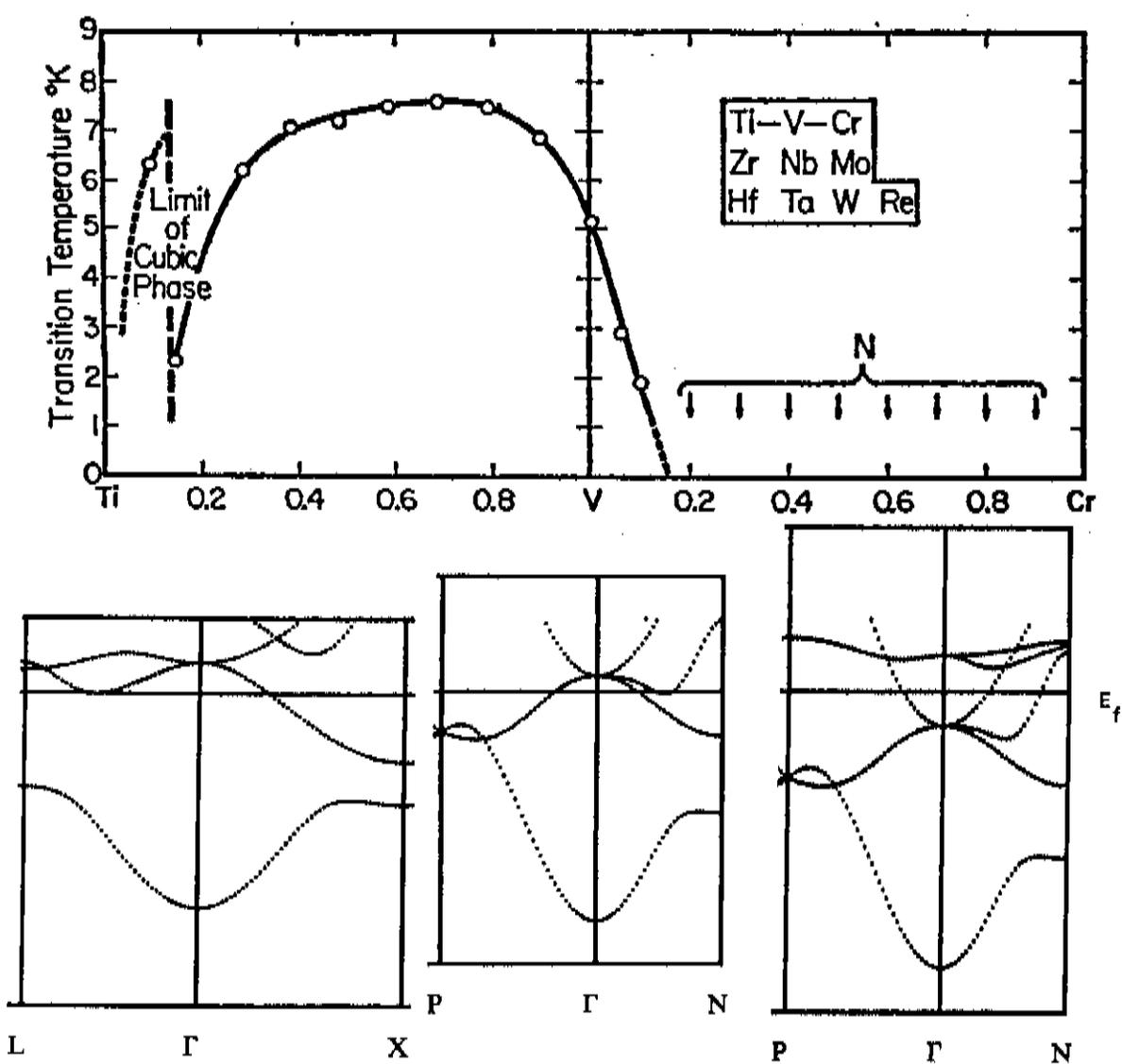

Figure 10: Dependence of $T_c$ on electron concentration for transition elements in the fourth row of the periodic table (from Ref. [58]). The band structures for Ti, V and Cr (from left to right) (from Ref. (50)) around the $\Gamma$ point are shown also.

The vast majority of workers in the field argue that $MgB_2$ is a conventional electron-phonon driven superconductor. The arguments used to support this view are that it exhibits an isotope effect, and that its properties look like conventional s-wave superconductivity. Both arguments are clearly open to question: concerning the first one, most high $T_c$ cuprates exhibit an isotope effect[63], and even so there is general agreement that they are not electron-phonon driven; furthermore, some conventional materials don't exhibit an isotope effect (e.g. Ru, Zr). Concerning the second one, suffice to say that s-wave superconductivity is not a signature of electron-phonon interaction driven superconductivity.

The theory of hole superconductivity provides clear predictions for the dependence of $T_c$ on doping $MgB_2$[5]. In contrast, electron-phonon theory has not provided clear predictions so far (yet undoubtedly will provide clear postdictions after experimental results are reported). We expect that a small amount of electron doping should increase $T_c$, as seen in Fig. 6. This is in apparent contradiction with experiments where Al is substituted for Mg[64]. We believe this may be due to sample quality or extrinsic effects, and that future experiments will show an increase in $T_c$ with electron doping.

More generally, and in contrast to electron-phonon theory, the theory of hole superconductivity provides clear guidelines in the search for higher superconducting transition temperatures: high $T_c$ will occur in materials where conduction occurs through holes in nearly filled bands, where the conduction occurs through negatively charged ions, and where the distance between ions is small and the connectivity is high. Unfortunately, these conditions are mutually conflicting, and they conflict with lattice stability. Hole states at the Fermi level imply many antibonding states in the band are occupied, leading to lattice instabilities; negatively charged ions will repel each other and not tend to form dense conducting substructures. Lattice instabilities will often come in and prevent the ideal structure and band filling to be realized. To the extent that these requirements can be met by clever material design or serendipity, higher $T_c$'s will be found, but for these reasons the search for higher $T_c$ materials will remain an elusive task, and require multi-atom compounds and elaborate structures. However, the fact that $T_c$ as high as 40K has been achieved in a compound as simple as $MgB_2$ suggests that room temperature superconductivity may just be around the corner.

The BCS-Migdal-Eliashberg theory of superconductivity provided a unique and seemingly consistent description of superconductivity for all known materials at the time reference [65] was written (1969), supposedly the 'last nail in the coffin of superconductivity' ([65], preface). Up to that time, when some materials showed anomalies they were dismissed or explained by additional assumptions[51]. Since then however, many materials have been discovered that clearly don't fit in the conventional picture: heavy fermion superconductors, high $T_c$ cuprates, strontium ruthenate, $Ba_{1-x}K_xBiO_3$, organics, alkali-doped $C_{60}$, electric-field doped $C_{60}$, and now $MgB_2$. Either $T_c$ is too high, phonon structure in tunneling is too small, or other properties do not fit the conventional framework. The effort to explain all superconductors by the conventional theory was hence abandoned long ago, and many new theories have been proposed to explain each of these new anomalous cases. It is certainly possible that there are many different mechanisms and manifestations of superconductivity in nature. It used to be also thought possible that with sufficient number of epicycles a consistent description of the motion of all celestial bodies could be achieved, with the Earth at the center[66]. Then, however, a simpler solution was found[67]. Similarly, we argue that the theory of hole superconductivity provides a simple and unique solution to explain all manifestations of superconductivity in materials.

**ACKNOWLEDGEMENT**

I am grateful to F. Marsiglio for collaboration in much of the work reviewed here.